\documentclass[useAMS,a4paper,fleqn,usenatbib]{mnras}
\usepackage{color}
\usepackage{graphicx}	
\usepackage{amsmath}	
\usepackage{amssymb}	
\usepackage{multicol}        
\usepackage{bm}		
\usepackage{pdflscape}	

\usepackage[T1]{fontenc}
\usepackage{ae,aecompl}

\newcommand{\qc}{NGC~4593}

\newcommand{\xmm}{{\em XMM-Newton}}
\newcommand{\nus}{{\em NuSTAR}}
\newcommand{\chandra}{{\em Chandra}}
\newcommand{\suz}{{\em Suzaku}}
\newcommand{\integral}{{\em INTEGRAL}}
\newcommand{\sax}{{\em BeppoSAX}}

\newcommand{\aer}[3]{$#1^{+ #2}_{- #3}$}
\newcommand{\aerm}[3]{#1^{+ #2}_{- #3}}

\newcommand{\ser}[2]{$#1 \pm #2$}
\newcommand{\serm}[2]{#1 \pm #2}
\newcommand{\serexp}[3]{($#1 \pm #2) \times 10^{#3}$}
\newcommand{\lsup}[1]{$< #1$}
\newcommand{\linf}[1]{$> #1$}

\newcommand{\expo}[2]{$ #1 \times 10^{#2}$}

\newcommand{\pers}{s$^{-1}$}
\newcommand{\fluxcgs}{ergs~s$^{-1}$~cm$^{-2}$}
\newcommand{\lumcgs}{ergs~s$^{-1}$}
\newcommand{\kms}{km~s$^{-1}$}
\newcommand{\sqcm}{cm$^{-2}$}

\newcommand{\chisq}{\chi^{2}}
\newcommand{\rchisq}{\chi^{2}/\textrm{dof}}
\newcommand{\dchi}{\Delta \chi^{2}}
\newcommand{\ddof}{\Delta \textrm{dof}}
\newcommand{\dcash}{\Delta \textrm{C}}
\newcommand{\rcash}{\textrm{C/dof}}

\newcommand{\cut}{E_{\textrm{c}}}
\newcommand{\nh}{N_{\textrm{H}}}

\newcommand{\nhone}{N_{\textrm{H,}\textsc{wa1}}}
\newcommand{\nhtwo}{N_{\textrm{H,}\textsc{wa2}}}
\newcommand{\xione}{\xi_{\textsc{wa1}}}
\newcommand{\xitwo}{\xi_{\textsc{wa2}}}
\newcommand{\vturbone}{\sigma_{v,\textsc{wa1}}}
\newcommand{\vturbtwo}{\sigma_{v,\textsc{wa2}}}
\newcommand{\vone}{v_{\textsc{wa1}}}
\newcommand{\vtwo}{v_{\textsc{wa2}}}
\newcommand{\afe}{A_{\textrm{Fe}}}

\newcommand{\rin}{R_{\textrm{in}}}
\newcommand{\rg}{$R_{\textrm{G}}$}
\newcommand{\rsxill}{$\mathcal{R}_{s,\textsc{xillver}}$}
\newcommand{\rfxill}{$\mathcal{R}_{f,\textsc{xillver}}$}
\newcommand{\rsrel}{$\mathcal{R}_{s,\textsc{relxill}}$}

\newcommand{\fek}{Fe~K~$\alpha$}

\newcommand{\ovii}{O~{\sc vii}}
\newcommand{\nvi}{N~{\sc vi}}

\newcommand{\xspec}{{\sc xspec}}
\newcommand{\xillver}{{\sc xillver}}
\newcommand{\relxill}{{\sc relxill}}

\newcommand{\cloudy}{{\sc cloudy}}
\newcommand{\diskbb}{{\sc diskbb}}

\usepackage[space]{grffile}
\usepackage[normalem]{ulem}
\usepackage{hyperref}
\usepackage{comment}
\bibpunct{(}{)}{;}{a}{}{,}

\begin{document}
\title[High-energy monitoring of NGC 4593]{High-energy monitoring of NGC~4593 with \textit{XMM-Newton} and \textit{NuSTAR}. X-ray spectral analysis}

\author[F. Ursini et al.]
	{F. Ursini,$^{1,2,3}$\thanks{e-mail: \href{mailto:francesco.ursini@univ-grenoble-alpes.fr}{\texttt{francesco.ursini@univ-grenoble-alpes.fr}}} 
	P.-O. Petrucci,$^{1,2}$
	G. Matt,$^{3}$ 
	S. Bianchi,$^{3}$ 
	M. Cappi,$^{4}$ \newauthor
	B. De Marco,$^{5}$ 
	A. De Rosa,$^{6}$
	J.~Malzac,$^{7,8}$ 
	A.~Marinucci,$^{3}$
	G. Ponti$^{5}$ 
	and
	A. Tortosa$^{3}$ \\
	$^1$ Univ. Grenoble Alpes, IPAG, F-38000 Grenoble, France. \\
	$^2$ CNRS, IPAG, F-38000 Grenoble, France. \\
	$^3$ Dipartimento di Matematica e Fisica, Universit\`a degli Studi Roma Tre, via della Vasca Navale 84, 00146 Roma, Italy. \\
	$^4$ INAF-IASF Bologna, Via Gobetti 101, I-40129 Bologna, Italy. \\
	$^5$ Max-Planck-Institut f\"ur extraterrestrische Physik, Giessenbachstrasse, D-85748 Garching, Germany. \\
	$^6$ INAF/Istituto di Astrofisica e Planetologia Spaziali, via Fosso del Cavaliere, 00133 Roma, Italy.\\
	$^7$ Universit\'e de Toulouse, UPS-OMP, IRAP, Toulouse, France.\\
	$^8$ CNRS, IRAP, 9 Av. colonel Roche, BP44346, F-31028 Toulouse cedex 4, France.
}

\date{Released Xxxx Xxxxx XX}


\maketitle

\label{firstpage}

\begin{abstract}
We present results from a joint \xmm/\nus\ monitoring of the Seyfert 1 NGC~4593, consisting of $5 \times 20$ ks simultaneous observations spaced by two days, performed in January 2015. The source is variable, both in flux and spectral shape, on time-scales down to a few ks and with a clear softer-when-brighter behaviour.
In agreement with past observations, we find the presence of a warm absorber well described by a two-phase ionized outflow.
The source exhibits a cold, narrow and constant \fek\ line at 6.4 keV, and a broad component is also detected.
The broad-band (0.3--79 keV) spectrum is well described by a primary power law with $\Gamma \simeq 1.6-1.8$ and an exponential cut-off varying from \aer{90}{40}{20} keV to \linf{700} keV, two distinct reflection components, and a variable soft excess correlated with the primary power law.
This campaign shows that probing the variability of Seyfert 1 galaxies on different time-scales is of prime importance to investigate the high-energy emission of AGNs.
\end{abstract}

\begin{keywords}
	galaxies: active --- galaxies: Seyfert --- X-rays: galaxies --- X-rays: individuals (NGC 4593)
\end{keywords}

\section{Introduction}
Active galactic nuclei (AGNs) are thought to be powered by an accretion disc around a supermassive black hole, mostly emitting in the optical/UV band. According to the standard paradigm, the X-ray emission is due to thermal Comptonization of the soft disc photons in a hot region, the so-called corona \cite[][]{haardt&maraschi1991,hmg1994,hmg1997}. This process explains the power-law shape of the observed X-ray spectrum of AGNs. A feature of thermal Comptonization is a high-energy cut-off, which has been observed around $\sim$ 100 keV in several sources thanks to past observations with \textit{CGRO}/OSSE \cite[][]{zdziarski2000}, \sax\ \cite{perola2002}, and more recently \textit{Swift}/BAT \cite[][]{bat70} and \integral\ \cite[][]{malizia2014,lubinski2016}. From the application of Comptonization models, such high-energy data allow to constrain the plasma temperature, which is commonly found to range from 50 to 100 keV \cite[e.g.,][]{zdziarski2000,lubinski2016}. Furthermore, the cut-off energy is now well constrained in an increasing number of sources thanks to the unprecedented sensitivity of \nus\ up to $\sim 80$ keV \cite[e.g.,][]{IC4329A_Brenneman,marinucci2014swift,balokovic2015,5548,ballantyne2014,matt20155506}.
The primary X-ray emission can be modified by different processes, such as absorption from neutral or ionized gas (the so-called warm absorber), and Compton reflection from the disc \cite[e.g.,][]{george&fabian1991,MPP1991} or from more distant material, like the molecular torus at pc scales \cite[e.g.,][]{matt2003}. A smooth rise below 1-2 keV above the extrapolated high-energy power law is commonly observed in the spectra of AGNs \cite[see, e.g.,][]{caixa1}. The origin of this so-called soft excess is uncertain \cite[see, e.g.,][]{done2012SE}. Ionized reflection is able to explain the soft excess in some sources \cite[e.g.,][]{crummy2006,ponti2006,walton2013}, while Comptonization in a ``warm'' region is favoured in other cases \cite[e.g.,][]{mehdipour2011509,rozenn2014mrk509SE}.

The current knowledge of the geometrical and physical properties of the X-ray corona is far from being complete, as in most sources we still lack good constraints on the coronal temperature, optical depth and geometry. The approach based on multiple, broad-band observations with a high signal-to-noise ratio is a powerful tool to study the high-energy emission in Seyfert galaxies. Strong and fast X-ray variability, both in flux and spectral shape, is a hallmark of AGNs, and of historical importance to rule out alternatives to supermassive black holes as their central engine \cite[e.g.,][]{elliot&shapiro1974}. The analysis of X-ray variability allows disentangling the different spectral components and constraining their characteristic parameters, as shown by recent campaigns on Mrk 509 \cite[][]{kaastra2011mrk509,pop2013mrk509} and NGC 5548 \cite[][]{kaastra2014science,5548}.

In this paper, we discuss results based on a joint \xmm\ and \nus\ monitoring program on \qc\ \cite[$z=0.009$,][]{z4593}, an X-ray bright Seyfert 1 galaxy hosting a supermassive black hole of $(9.8 \pm 2.1) \times 10^6$ solar masses \cite[from reverberation mapping; see][]{mbh4593}. This is the first monitoring carried out by \xmm\ and \nus\ specifically designed to study the high-energy emission of an AGN through the analysis of its variability on time-scales of days. We reported preliminary results from a phenomenological timing analysis in \cite{procs_4593}. Here we focus on the broad-band (0.3--80 keV) X-ray spectral analysis.
Past observations of \qc\ with \sax\ \citep{guainazzi1999}, \xmm\ \citep{reynolds2004,brenneman2007} and \suz\ \citep{markowitz2009} have shown a strong reflection bump above 10 keV 
and the presence of two narrow \fek\ emission lines at 6.4 and around 7 keV. 
A soft excess below 2 keV has been reported, both in the 2002 \xmm\ data \citep{brenneman2007} and in the 2007 \suz\ data, with a drop in the 0.4--2 keV flux by a factor $>20$ in the latter \citep{markowitz2009}. 
\cite{guainazzi1999} found a lower limit on the high-energy cut-off of 150 keV from \sax\ data.

This paper is organized as follows. In Sect. \ref{sec:obs}, we describe the observations and data reduction. In Sect. \ref{sec:analysis} we present the analysis of the \xmm\ and \nus\ spectra. In Sect. \ref{sec:discussion}, we discuss the results and summarize our conclusions.
	\begin{table}
		\begin{center}
			\caption{The logs of the joint \xmm\ and \nus\ observations of \qc. \label{obs}}
			\begin{tabular}{ c c c c c } 
				\hline \rule{0pt}{2.5ex} Obs. &  Satellites & Obs. Id. & Start time (UTC)  & Net exp.\\ 
				& & & yyyy-mm-dd & (ks)  \\ \hline \rule{0pt}{2.5ex}
				1 & \xmm & 0740920201&  2014-12-29  & 16\\ 
				& \nus & 60001149002 &  &  22\\ \hline \rule{0pt}{2.5ex}
				2 & \xmm & 0740920301 & 2014-12-31 &  17 \\
				& \nus & 60001149004 &  & 22\\ \hline \rule{0pt}{2.5ex}
				3 & \xmm & 0740920401 &    2015-01-02  & 17\\
				& \nus & 60001149006 &  & 21 \\ \hline \rule{0pt}{2.5ex}
				4 & \xmm & 0740920501& 2015-01-04  & 15\\ 
				& \nus & 60001149008 &   & 23 \\ \hline \rule{0pt}{2.5ex}
				5 & \xmm & 0740920601 & 2015-01-06  & 21 \\
				& \nus & 60001149010 &   & 21\\ \hline
			\end{tabular}
		\end{center}
	\end{table}
	\begin{figure*}
		\includegraphics[width=2\columnwidth]{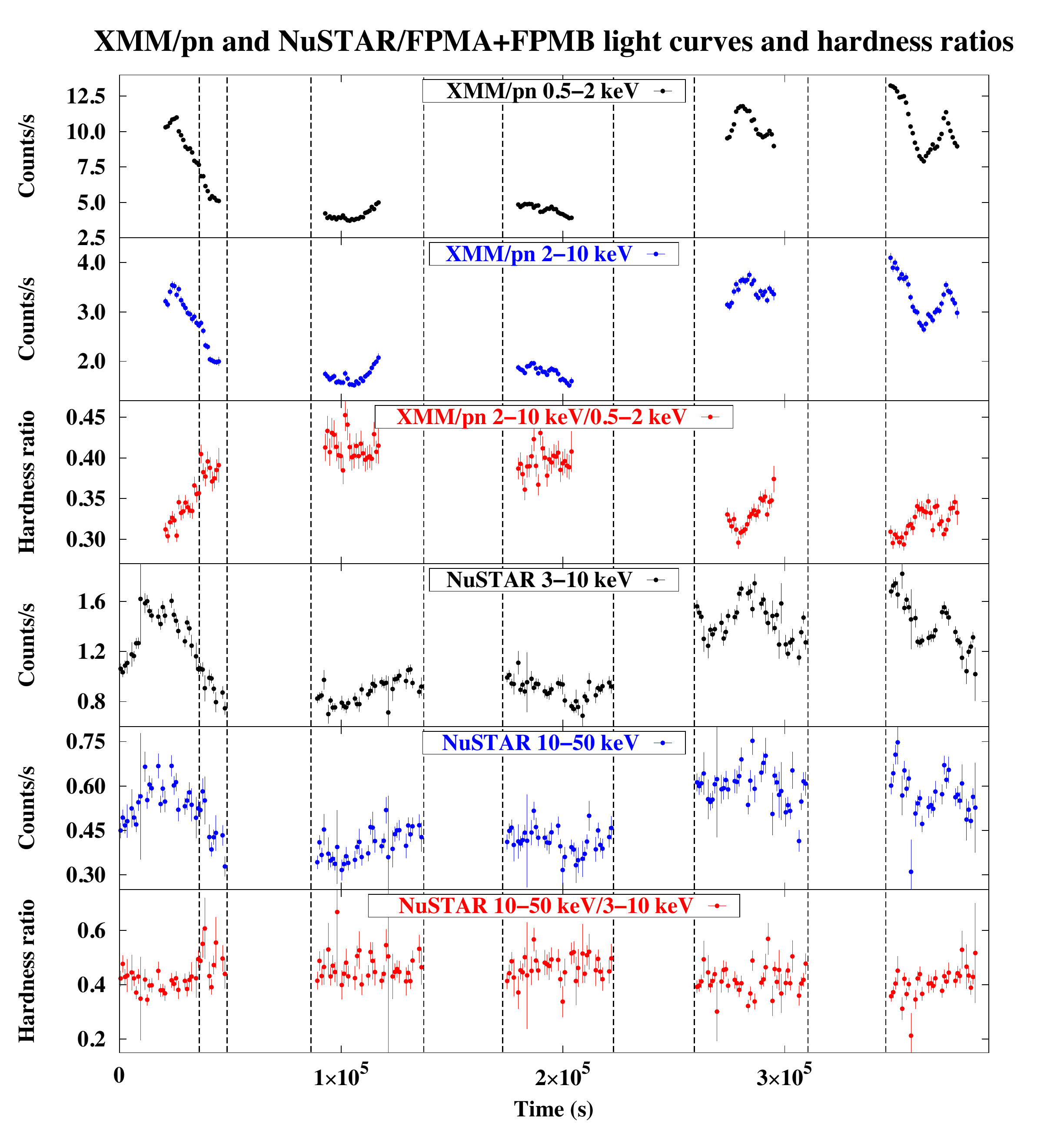}
		\caption{The light curves of the five joint \xmm\ and \nus\ observations of \qc. Time bins of 1 ks are used. Top panel: the \xmm/pn count-rate light curve in the 0.5--2 keV band. Second panel: the \xmm/pn count-rate light curve in the 2--10 keV band. Third panel: the \xmm/pn hardness ratio (2--10/0.5--2 keV) light curve. Fourth panel: the \nus\ count-rate light curve in the 3--10 keV band (FPMA and FPMB data are co-added). Fifth panel: the \nus\ count-rate light curve in the 10--50 keV band. Bottom panel: the \nus\ hardness ratio (10--50/3--10 keV) light curve. The vertical dashed lines separate the time intervals used for the spectral analysis. 
			\label{lc}}
	\end{figure*}
\section{Observations and data reduction}\label{sec:obs}
\xmm\ \cite[][]{xmm} and \nus\ \cite[][]{harrison2013nustar} simultaneously observed \qc\ five times, spaced by two days, between 29 December 2014 and 6 January 2015. Each pointing had an exposure of $\sim 20$ ks. We report in Table \ref{obs} the log of the data sets.

\xmm\ observed the source using the EPIC cameras \cite[][]{pn,MOS} and the Reflection Grating Spectrometer \cite[RGS;][]{RGS}. The EPIC instruments were operating in the Small Window mode, with the medium filter applied. Because of the significantly lower effective area of the MOS detectors, we only report the results obtained from pn data. 
The data were processed using the \xmm\ Science Analysis System (\textsc{sas} v14). Source extraction radii and screening for high-background intervals were determined through an iterative process that maximizes the signal-to-noise ratio \cite[][]{pico2004}. The background was extracted from circular regions with a radius of 50 arcsec, while the source extraction radii were in the range 20--40 arcsec. The EPIC-pn spectra were grouped such that each spectral bin contains at least 30 counts, and not to oversample the spectral resolution by a factor greater than 3. The RGS data were extracted using the standard \textsc{sas} task \verb|rgsproc|. 

The \nus\ data were reduced using the standard pipeline (\textsc{nupipeline}) in the \nus\ Data Analysis Software (\textsc{nustardas}, v1.3.1; part of the \textsc{heasoft} distribution as of version 6.14), using calibration files from \nus\ {\sc caldb} v20150316. Spectra and light curves were extracted using the standard tool {\sc nuproducts} for each of the two hard X-ray detectors aboard \nus, which sit inside the corresponding focal plane modules A and B (FPMA and FPMB). The source data were extracted from circular regions with a radius of 75 arcsec, and background was extracted from a blank area
with the same radius,
close to the source. The spectra were binned to have a signal-to-noise ratio greater than 5 in each spectral channel, and not to oversample the instrumental resolution by a factor greater than 2.5. The spectra from FPMA and FPMB were analysed jointly but not combined. 

In Fig. \ref{lc} we show the light curves in different energy ranges of the five \xmm\ and \nus\ observations. The light curves exhibit a strong flux variability, up to a factor of $\sim 2$, on time-scales as short as a few ks. In Fig. \ref{lc} we also show the \xmm/pn (2--10 keV)/(0.5--2 keV) hardness ratio light curve and the \nus\ (10--50 keV)/(3--10 keV) hardness ratio light curve. These light curves show a significant spectral variability, particularly in the softer band, with a less prominent variability in the harder band. From these plots, it can be seen that a higher flux is associated with a lower hardness ratio, i.e. we observe a ``softer when brighter'' behaviour \cite[see also][]{procs_4593}. During the first observation, both flux and spectral variability are particularly strong. To investigate the nature of these variations, we analysed the spectra of each observation separately, splitting only the first observation into two $\sim 10$ ks long bits (see Fig. \ref{lc}). We did not split the time intervals for the extraction of the spectra further, because in the other observations the spectral variability is less pronounced.
This choice leaves us with a total of six spectra from our campaign, which we analyse in the following.
\section{Spectral analysis}\label{sec:analysis}
Spectral analysis and model fitting was carried out with the \xspec\ 12.8 package \cite[][]{arnaud1996}. RGS spectra were not binned and were analysed using the C-statistic \cite[][]{cstat}, to take advantage of the high spectral resolution of the gratings. Broad-band (0.3--80 keV) fits were instead performed on the binned EPIC-pn and \nus\ spectra, using the $\chisq$ minimisation technique. All errors are quoted at the 90\% confidence level ($\dcash = 2.71$ or $\dchi = 2.71$) for one interesting parameter.

\begin{figure}
	\includegraphics[width=\columnwidth]{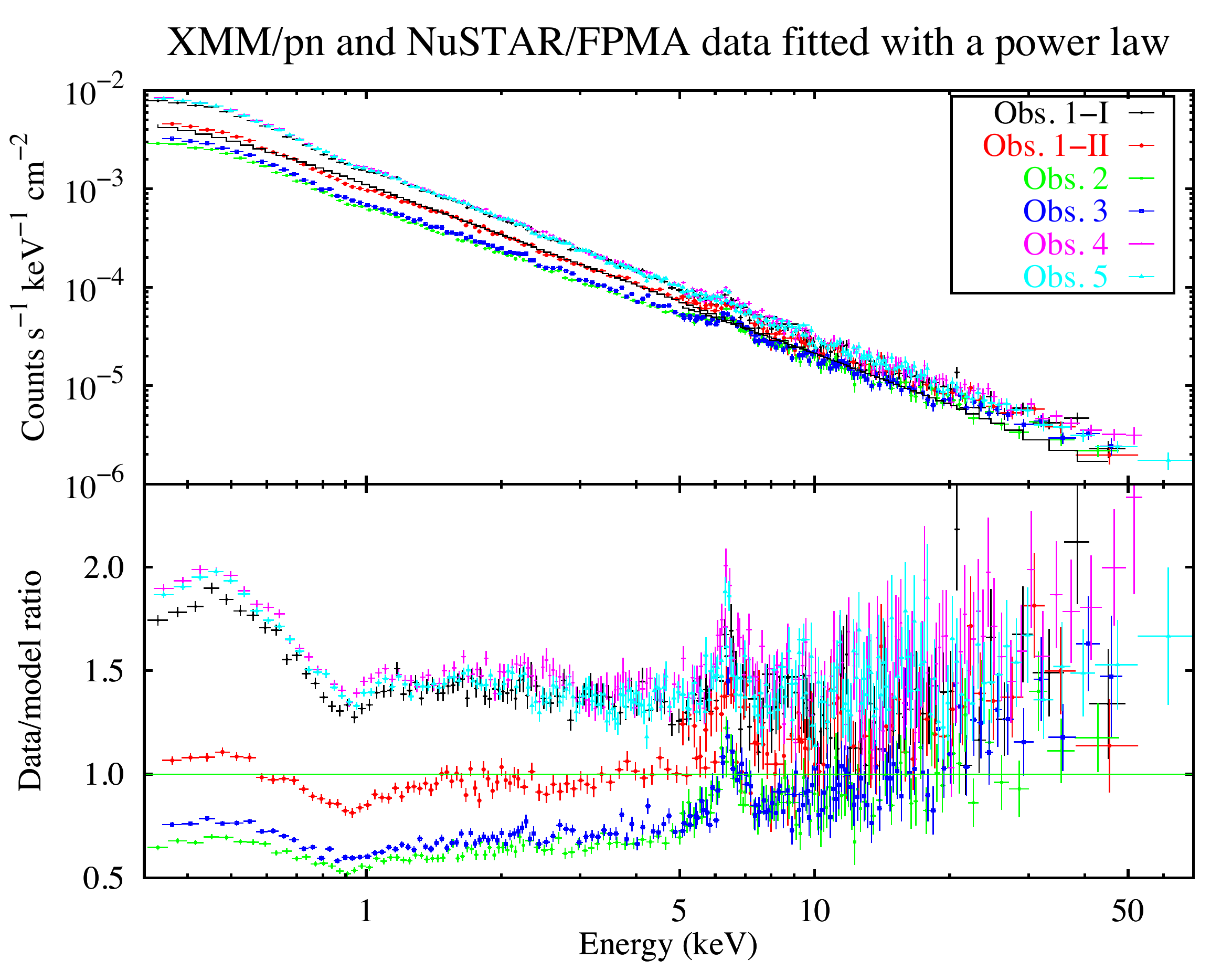}
	\caption{Upper panel: the six \xmm/pn and \nus~spectra of \qc. Lower panel: the ratio of the six spectra to a single power law. Only \nus/FPMA data are shown for clarity.
		\label{spectra}}
\end{figure}

In Fig. \ref{spectra} we plot the six \xmm/pn and \nus/FPMA spectra, simultaneously fitted with a power law with tied parameters for comparison. The plot in Fig. \ref{spectra} shows that the spectral variability is especially strong in the soft band, i.e. below 10 keV. Furthermore, the ratio of the spectra to the power law below 1 keV suggests the presence of a soft excess which is stronger in the high flux states. The prominent feature at 6-7 keV can be attributed to the \fek\ emission line at 6.4 keV \cite[][]{reynolds2004,brenneman2007,markowitz2009}, as we will discuss in the following.

To fully exploit the data set, we always fitted all the \xmm/pn and \nus/FPMA and FPMB data simultaneously, allowing for free cross-calibration constants. The FPMA and FPMB modules are in good agreement with pn, with cross-calibration factors of $1.02 \pm 0.01$ and $1.05 \pm 0.01$ respectively, fixing the constant for the pn data to unity. All spectral fits include neutral absorption ({\sc phabs} model in {\sc xspec}) from Galactic hydrogen with column density $\nh = 1.89 \times 10^{22}$ \sqcm\ \cite[][]{kalberla2005}. We assumed the element abundances of \cite{lodders2003}.
\subsection{The RGS spectra}\label{subsec:rgs}
\begin{figure}
	\includegraphics[width=\columnwidth]{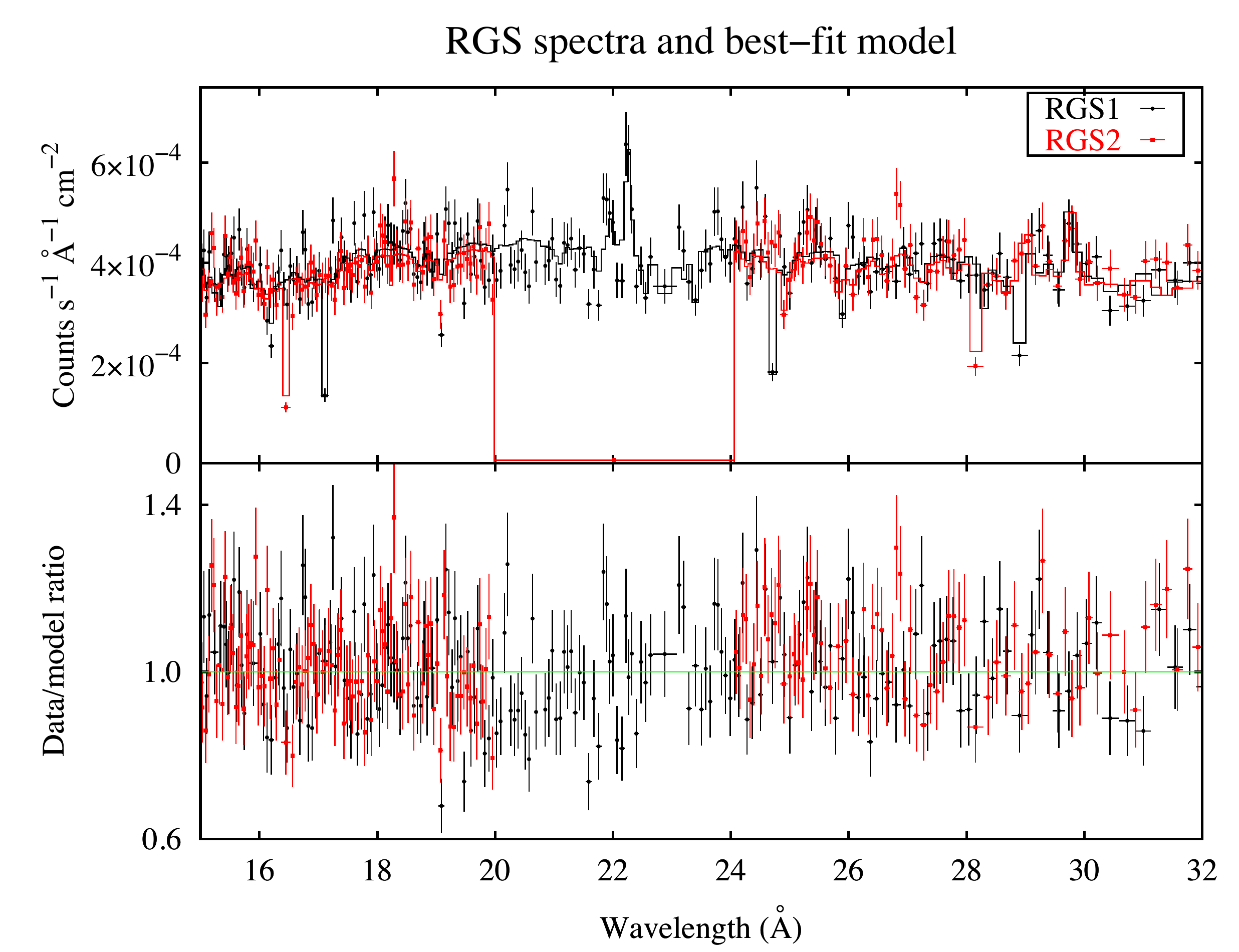}
	\caption{Upper panel: the \xmm/RGS spectra (15--32 \AA) with the best-fitting model. Data are rebinned for displaying purposes only. Lower panel: the ratio of the spectra to the model. 
		\label{rgsfit}}
\end{figure}

\qc\ is known to host a warm absorber from previous observations with \chandra\ and \xmm\ \cite[][]{mckernan_4593,steenbrugge_4593,brenneman2007,ebrero_4593}. The location of this ionized absorber is at least a few pc from the central source \cite[][]{ebrero_4593}, therefore we do not expect it to significantly vary on a few days time-scale. We thus co-added the RGS data from the different epochs, for each detector, to obtain a better signal-to-noise ratio.
We fitted the RGS1 and RGS2 co-added spectra in the 0.3--2 keV band. We used a simple model including a power law and the warm absorber. To find a good fit, we needed to include two ionized absorbers, both modelled using the spectral synthesis code \cloudy\ \cite[][]{cloudy}. We built a \cloudy\ table model with an allowed range for the ionization parameter $\log \xi$ of 0.1--4.9 (in units of \lumcgs\ cm), while the allowed range for the column density is $10^{19}$--$10^{24}$ \sqcm. These ranges are suitable for both absorption components. Starting with a fit including only one absorption component, we find $\rcash=6131/5388$. Adding a second component, we find $\rcash = 5941/5384$ ($\dcash/\ddof=-190/-4$) and positive residuals around 22 and 29 \AA, that may be attributed to the K~$\alpha$ triplets of the He-like ions \ovii\ and \nvi. To constrain the parameters of such emission lines, we performed two local fits at the corresponding wavelengths, on intervals $\sim 100$ channels wide. Due to the small bandwidth, the underlying continuum is not sensitive to variations of the photon index. We thus fixed it at 2, leaving free the normalization of the power law. With such fits, we detect four significant (at $90 \%$ confidence level) emission lines.
We identify the lines as the intercombination ($1s^2 \,^1S_0 - 1s 2p \,^3P_{2,1}$) and forbidden ($1s^2 \,^1S_0 - 1s 2s \,^3S_1$) components of the \ovii\ K~$\alpha$ triplet, and the resonance ($1s^2 \,^1S_0 - 1s 2p \,^1P_1$) and forbidden components of the \nvi\ K~$\alpha$ triplet (see Table \ref{lines}). 
We first modelled these lines using four Gaussian components. After including the lines in the fit over the whole energy range, we find $\rcash = 5869/5376$ (i.e. $\dcash / \ddof = -72/{-}8$). 
Then, we replaced the Gaussian lines with the emission spectrum from a photoionized plasma, modelled with a \cloudy\ table having the ionization parameter, the column density and the normalization as free parameters. We find a nearly equivalent fit, $\rcash = 5866/5379$ ($\dcash / \ddof = -3/{+}3$) and no prominent residuals that can be attributed to strong atomic transitions (see Fig. \ref{rgsfit}). 
The best-fitting parameters of the warm absorbers and of the emitter are reported in Table \ref{rgs}. 
\begin{table}
	\begin{center}
		\caption{The emission lines detected in RGS spectra. $\lambda_T$ and $E_T$ are the theoretical wavelength and energy of the lines (rest-frame), as reported in the {\sc atomdb} database \citep{atomdb}. $\sigma$ is the intrinsic line width and $v$ is the velocity shift with respect to the systemic velocity of \qc; these two parameters were fixed between lines of the same multiplet. The flux is in units of $10^{-5}$ photons \sqcm\ \pers. \label{lines}}
		\begin{tabular}{l c c c c c}
			\hline \rule{0pt}{2.5ex}
			line id. & $\lambda_T$ (\AA) & $E_T$ (keV) & $\sigma$ (eV) & $v$ (\kms) & flux \\ \hline \rule{0pt}{2.5ex}
			\ovii\ (f) & 22.091 & 0.561 & \aer{ 0.7 }{ 0.4 }{ 0.6 }  & \ser{ -450 }{ 110 }  & \aer{ 6.2 }{ 2.0 }{1.9} \\ \rule{0pt}{2.5ex} \ovii\ (i) & 21.795 & 0.569 & - & - &  \aer{ 2.3 }{ 1.8 }{ 1.6 } \\ \hline \rule{0pt}{2.5ex}
			\nvi\ (r) & 28.738 & 0.431 & \lsup{ 1 }  & \aer{ -360 }{270}{180}  & \aer{ 4.0 }{ 1.4 }{3.2} \\ \rule{0pt}{2.5ex} \nvi\ (f) & 29.528 & 0.420 & - & - &  \aer{ 4.6 }{ 2.7 }{ 2.1 } \\ \hline
		\end{tabular}
	\end{center}
\end{table}
\begin{table}
	\begin{center}
		\caption{Best-fitting parameters of the warm absorbers 
			(\textsc{wa1, wa2}) and photoionized emission (\textsc{em}), for the RGS spectra. 
			For all these models, $\xi$ is the ionization parameter,
			$\sigma_v$ is the turbulent velocity, $\nh$ is the column density and $v$ is the velocity shift with respect to the systemic velocity of \qc; $N_{\textsc{em}}$ is the normalization of the photoionized emission table.
		\label{rgs}
		}
		\begin{tabular}{l c}
			\hline \rule{0pt}{2.5ex}
			$\log \xione $ (\lumcgs\ cm) & \ser{ 2.22 }{ 0.01 } 
			\\  \rule{0pt}{2.5ex} $\log \vturbone$ (\kms) & \ser{ 1.72 }{ 0.07 }
			\\  \rule{0pt}{2.5ex} $\log \nhone$ (\sqcm) & \aer{ 21.83 }{ 0.07 }{ 0.02 } 
			\\  \rule{0pt}{2.5ex}  $\vone$ (\kms) & \ser{ -870 }{ 60 } 
			\\  \hline \rule{0pt}{2.5ex} 
			$\log \xitwo$ (\lumcgs\ cm) & \ser{ 0.43 }{ 0.08 }
			\\ \rule{0pt}{2.5ex} $\log \vturbtwo$ (\kms) & unconstr.
			\\ \rule{0pt}{2.5ex} $\log \nhtwo$ (\sqcm)& \ser{ 20.90 }{ 0.02 } 
			\\ \rule{0pt}{2.5ex} $\vtwo$ (\kms)& \ser{ -300 }{ 200 } 
			\\ \hline \rule{0pt}{2.5ex} 
			$\log \xi_{\textsc{em}}$ (\lumcgs\ cm) & \ser{ 0.56 }{ 0.03 } 
			\\ \rule{0pt}{2.5ex} $\log \sigma_{v,\textsc{em}}$ (\kms) & unconstr. 
			\\ \rule{0pt}{2.5ex} $\log N_{\textrm{H},\textsc{em}}$ (\sqcm) & \linf{21.4} 
			\\ \rule{0pt}{2.5ex} $v_{\textsc{em}}$ (\kms) & \ser{ -720 }{ 60 }
			\\ \rule{0pt}{2.5ex} $N_{\textsc{em}}$ ($\times 10^{-18}$) & \ser{ 6 }{1} \\ \hline
		\end{tabular}
	\end{center}
\end{table}
\subsection{The Fe K~$\alpha$ line}\label{subsec:line}
We next investigated the properties of the \fek\ line at 6.4 keV.
We focused on \xmm/pn data, given their better energy resolution compared with \nus\ in that energy band. We fitted the six pn spectra between 3 and 10 keV, to avoid the spectral complexities in the soft X-ray band. 
The model included a power law plus a Gaussian line at 6.4 keV. Due to a known calibration problem in pn data, the energy of the Gaussian line is blueshifted by $\sim 50$ eV with respect to the theoretical value of 6.4 keV \cite[for a detailed discussion, see][]{cappi_5548_arxiv}. The shifts vary between different observations and are present using either single plus double or single-only events, despite the use of the latest correction files for charge transfer inefficiency (CTI) and procedure as described in Smith et al. (2014, \textsc{xmm-ccf-rel-323}\footnote{\url{http://xmm2.esac.esa.int/docs/documents/CAL-SRN-0323-1-1.ps.gz}}). This is likely an effect of the long-term degradation of the EPIC/pn CTI. To correct for such uncertainty, we fixed the Gaussian line energy at 6.4 keV, implying production by ``cold'' iron, i.e. less ionized than Fe \textsc{xii}, while leaving the redshift free to vary in pn data. We find an acceptable fit ($\rchisq= 648/571$) assuming an intrinsically narrow line, i.e. its intrinsic width $\sigma$ is fixed at zero. If $\sigma$ is left free, we find a slightly better fit ($\rchisq= 643/570$, i.e. $\dchi/\ddof = -5/{-}1$), with $\sigma=50 \pm 30$ eV and a probability of chance improvement (calculated with the $F$-test) around $0.04$.  

\begin{figure}
	\includegraphics[width=\columnwidth]{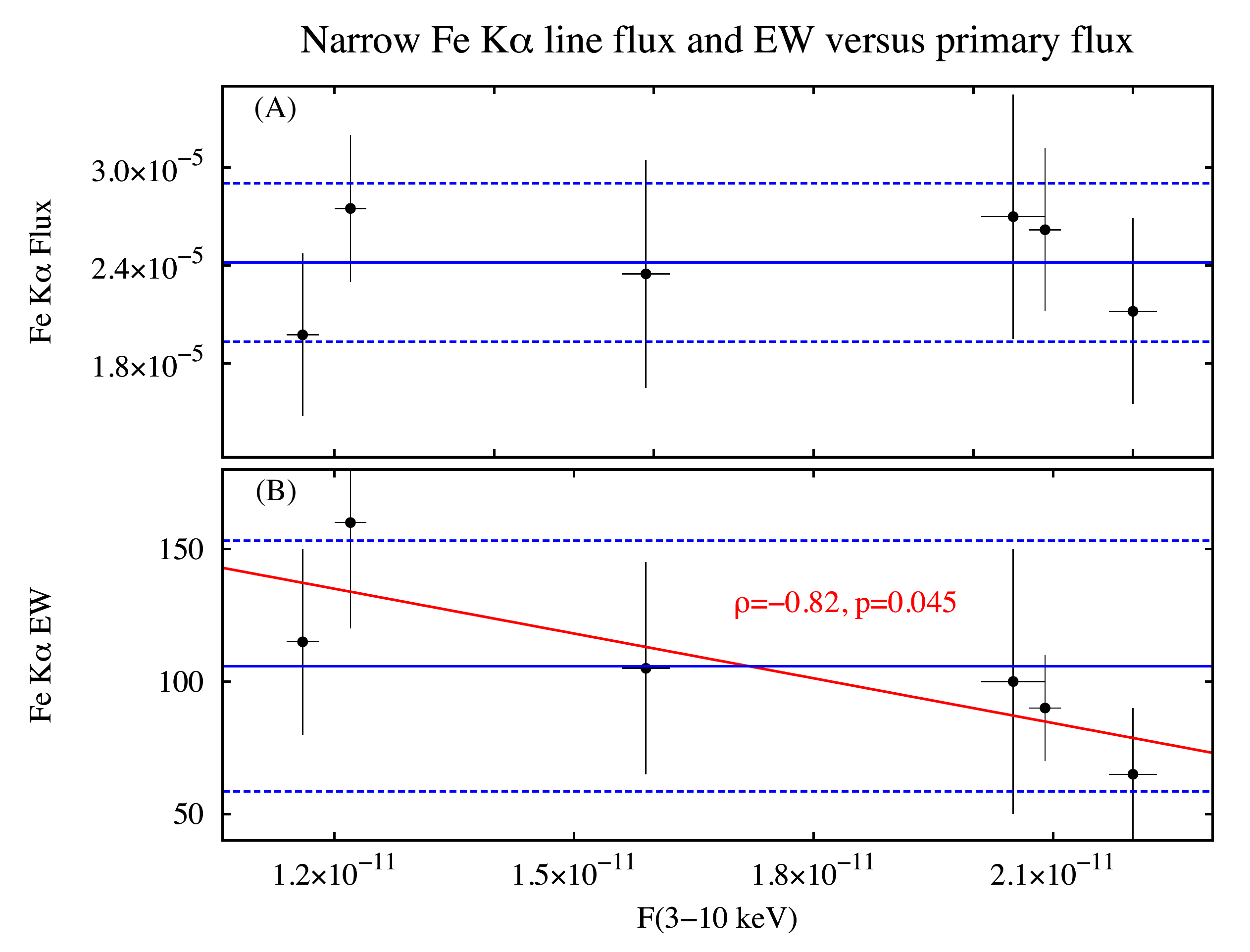}
	\caption{Parameters of the narrow \fek\ line at 6.4 keV, plotted against the primary flux in the 3--10 keV band.
		Panel (A): the line flux in units of 
		photons \sqcm~\pers. 
		Panel (B): the line equivalent width (EW) in units of eV. 
		Error bars denote the 1-$\sigma$ uncertainty. The blue solid lines represent the mean value for each parameter, while the blue dashed lines represent the standard deviation (i.e. the root mean square of the deviations from the mean). The red line represents a linear fit to the data. }
	\label{cor1}
\end{figure}

\begin{figure}
	\includegraphics[width=\columnwidth]{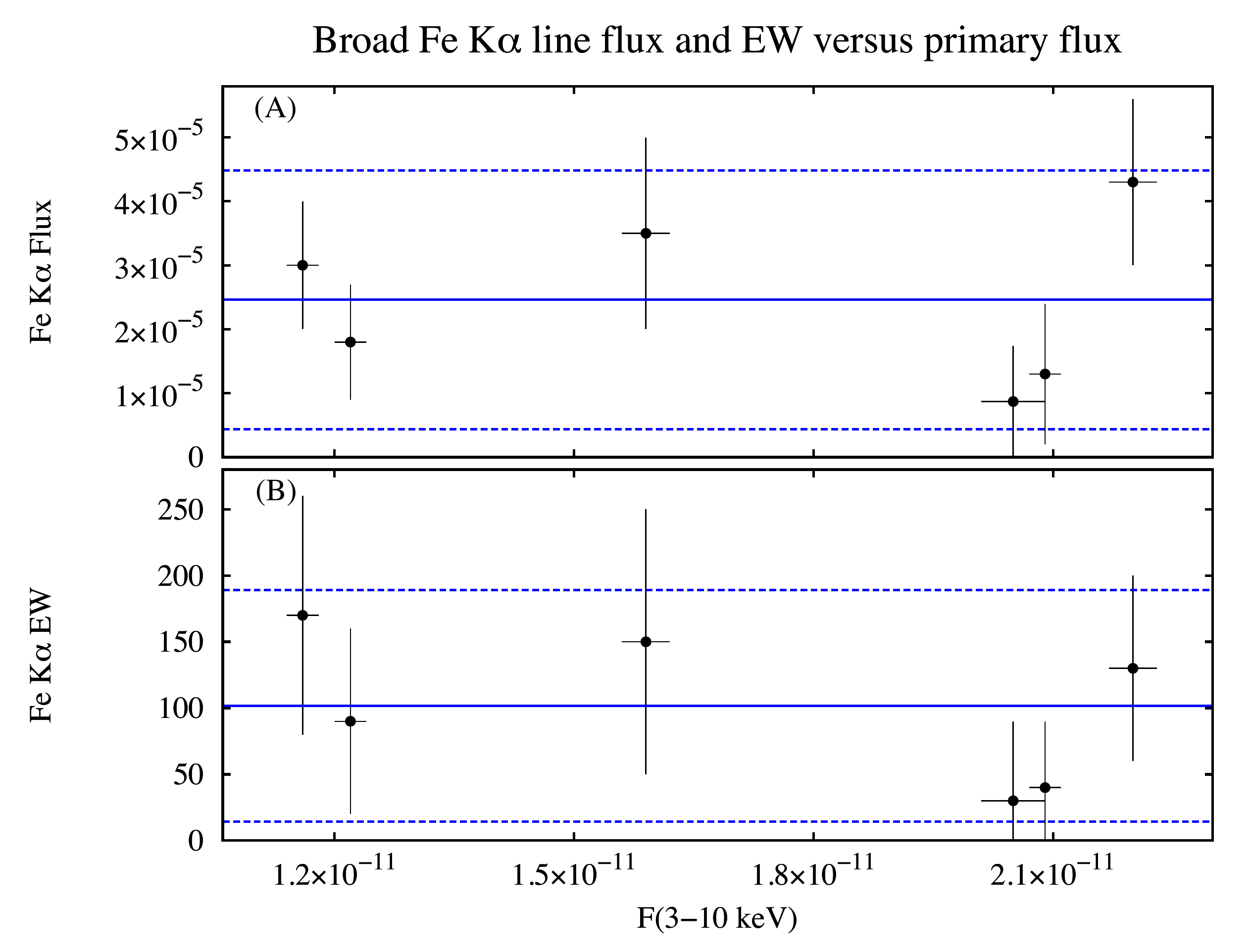}
	\caption{Parameters of the broad \fek\ line at 6.4 keV, plotted against the primary flux in the 3--10 keV band.
		Panel (A): the line flux in units of 
		photons \sqcm~\pers. 
		Panel (B): the line equivalent width (EW) in units of eV. 
		Error bars denote the 1-$\sigma$ uncertainty. The blue solid lines represent the mean value for each parameter, while the blue dashed lines represent the standard deviation. }
	\label{cor2}
\end{figure}

We then looked for the presence of a broad line component at 6.4 keV.
We added a second Gaussian line at 6.4 keV, leaving free the intrinsic width of both lines. The fit improves significantly ($\rchisq= 608/563$, i.e. $\dchi/\ddof = -35/{-}7$). The narrow line component has now an intrinsic width consistent with zero, therefore we fixed it at $\sigma=0$. 
The broad component has $\sigma=\aerm{450}{110}{70}$ eV.
We report in Fig. \ref{cor1} and \ref{cor2} the flux and equivalent width of both the narrow and broad components, plotted against the primary flux in the 3--10 keV band. 
The strong variability of the primary flux is not accompanied by a corresponding variability of the narrow line flux, which is consistent with being constant. As a consequence, the equivalent width is anticorrelated with the primary flux. The Pearson's coefficient is $-0.82$, with a $p$ value of \expo{4.5}{-2}.
The broad line flux, instead, shows hints of variability (a fit with a constant gives a reduced chi-square of 1.44), while the equivalent width is consistent with being constant (a fit with a constant gives a reduced chi-square of 0.67; the Pearson's coefficient is $-0.55$, with a $p$ value of 0.25). 
However, by including only two Gaussian lines at 6.4 keV, we find positive residuals around 7 keV which could be a potential signature of line emission from ionized iron, such as the K~$\alpha$ line of Fe \textsc{xxvi} at 6.966 keV (rest-frame), or of the K~$\beta$ line of neutral iron at 7.056 keV. \cite{brenneman2007} found evidence for a narrow, hydrogen-like \fek\ emission line in this source by using \xmm/pn data. \cite{markowitz2009} found only an upper limit on the presence of this line by using \suz\ data, but instead found a significant neutral Fe K~$\beta$ line. To test for the presence of such components, we included a third, narrow Gaussian line. 
We kept the line flux tied between different observations. Fixing the line energy at 6.966 keV, we find $ \rchisq = 596/573$ ($\dchi/\ddof = -20 / {-}1$), and no improvement by adding another Gaussian line at 7.056 keV. The fit is only slightly worse if we fix the energy of the third line at 7.056 keV instead of 6.966 keV ($\dchi = +2$). 
We measure a flux of \serexp{6}{2}{-6} photons \sqcm\ \pers\ for this line, and an equivalent width ranging from \ser{25}{10} eV to \ser{40}{20} eV. The ratio of the flux to that of the narrow \fek\ line is \ser{0.25}{0.14}, roughly consistent with the predicted ratio of fluorescence yields for K~$\beta$/K~$\alpha$ \cite[$\sim 0.13$, see][]{kaastra&mewe1993,molendi2003}.
Finally, we note that the intrinsic width of the broad component at 6.4 keV is now found to be $\sigma=\aerm{300}{130}{70}$ eV, while all the other parameters of the narrow and broad components are not significantly altered by the addition of the third line.
We also tried to thaw the energy of the broad component, to test if it could be due to ionized material. However, the fit is not significantly improved, and the line energy is found to be $E=\serm{6.44}{0.11}$ keV.

\begin{figure}
	\includegraphics[width=\columnwidth]{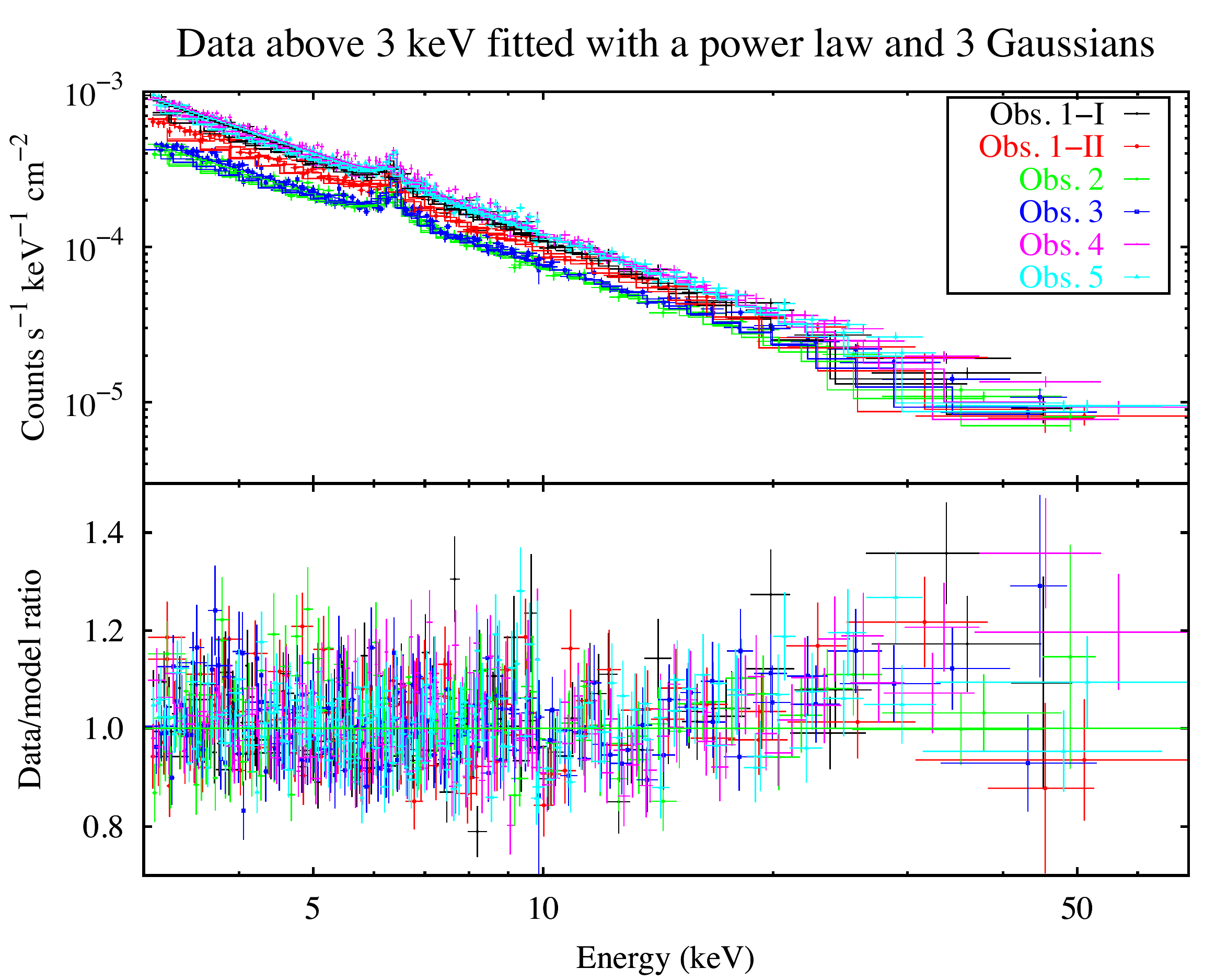}
	\caption{Upper panel: the pn and \nus\ spectra above 3 keV fitted with a model including a simple power law and three Gaussian lines. Lower panel: the ratio of the spectra to the model, showing an excess around 30 keV. The data were binned for plotting purposes.}
	\label{curv2}
\end{figure}

Overall, these results are consistent with the cold, narrow \fek\ line being produced by reflection off relatively distant material, lying at least a few light days away from the nucleus. A statistically significant broad component is detected, showing hints of variability, albeit with no clear trend with respect to the primary flux.
The results on the iron lines are summarized in Table \ref{iron}.
Finally, we had a first look at the presence of reflection components associated with the iron lines, using the energy range from 3 keV up to 79 keV. We fitted the \xmm/pn and \nus\ data with a simple power law plus three Gaussian emission lines, i.e. the narrow lines at 6.4 keV and 6.966 keV and the broad component at 6.4 keV, finding $\rchisq = 2084/1923$. The residuals indicate the presence of a moderate Compton hump peaking at around 30 keV (see Fig. \ref{curv2}).
\subsection{The broad-band fit}\label{subsec:broad}
After getting a clue to the origin of the iron line and to the presence of a reflection component, we fitted the pn and \nus\ data in the whole X-ray energy range (0.3--79 keV). We modelled the primary continuum with a cut-off power law. The other components are described below. 
In \xspec\ notation, the model reads: \textsc{phabs*WA1*WA2*(soft lines + cutoffpl + diskbb + relxill + xillver)}, where \textsc{WA1}, \textsc{WA2} and \textsc{soft lines} denote the \cloudy\ tables described in Sec. \ref{subsec:rgs}.
\paragraph*{The reflection component:} In Sect. \ref{subsec:line}, we found the presence of both a narrow and a broad \fek\ line components together with a Compton hump. We thus modelled the reflection component with \xillver\ \cite[][]{xillver1,xillver2} and \relxill\ \cite[][]{relxill} in \xspec. Both models self-consistently incorporate fluorescence lines and the Compton reflection hump, and in \relxill\ such components are relativistically blurred. We used the {\sc xillver-a-ec4} version, which takes into account the angular dependence of the emitted radiation and includes a primary power law with an exponential cut-off up to 1 MeV\footnote{\url{http://hea-www.cfa.harvard.edu/~javier/xillver/}}, and the \relxill\ version 0.4c\footnote{\url{http://www.sternwarte.uni-erlangen.de/~dauser/research/relxill/index.html}}.
Since the inclination angle of both reflectors was poorly constrained, we fixed it to 50 deg, consistently with past studies on this source \cite[][]{nandra1997,guainazzi1999}. 
We assumed for \xillver\ and \relxill\ the same iron abundance $\afe$, which was a free parameter of the fit, but tied between the different observations.

In \relxill, the normalization was left free to vary between the different observations. The input photon index and cut-off energy were tied to that of the primary power law, as this reflection component was assumed to originate relatively close to the primary source. \textit{A priori}, also the ionization could change among the observations, responding to the different ionizing flux from the primary source. However, the ionization parameter was always found to be consistent with a common value of $\log \xi \simeq 3$, therefore we tied this parameter between the different observations. Similarly, the inner radius of the disc was a free parameter of the fit, but tied between the different observations since it was consistent with being constant. Since the emissivity index was poorly constrained, we fixed it at 3 \cite[the classical case, see e.g.][]{wilkins&fabian2012}. We found no improvement by assuming a broken power law for the emissivity profile. Moreover, since the black hole spin parameter $a$ was poorly constrained, we assumed $a=0.998$. 

The \xillver\ component was consistent with being constant, being associated with the narrow, constant component of the \fek\ line. The normalization was a free parameter of the fit, but tied between the different observations. Since the input photon index and cut-off energy of \xillver\ were poorly constrained, we fixed them to the average values of the photon indexes and of the cut-off energies of the primary power law found in each observation.
We fixed the ionization parameter at the minimum value in \xillver, $\log \xi = 0$, with no improvement by leaving it free. 

\begin{figure*}
	\includegraphics[width=2\columnwidth]{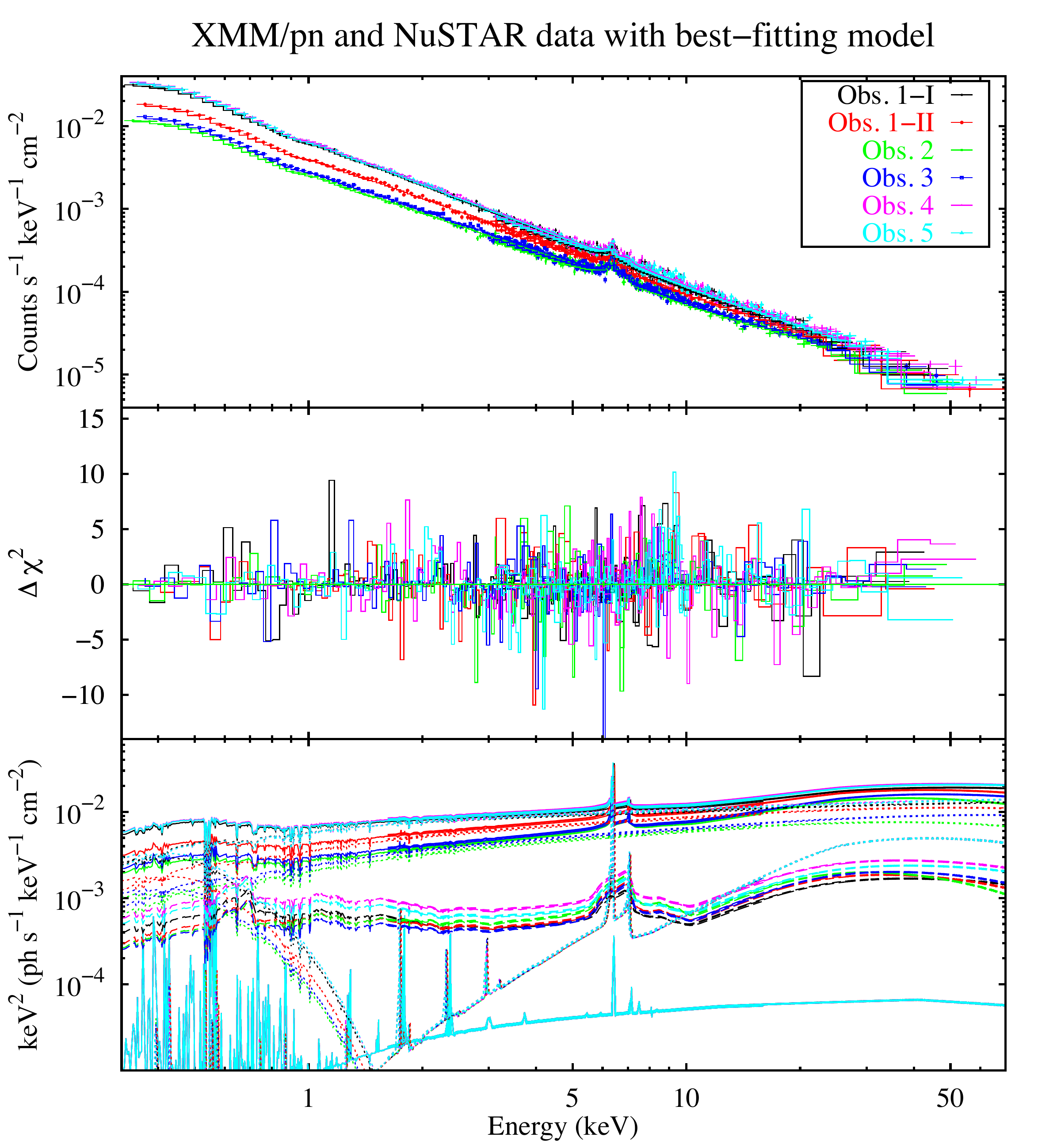}
	\caption{Broad-band X-ray data and best-fitting model (see Table \ref{params}). Upper panel: \xmm/pn and \nus\ data and folded model. Middle panel: contribution to $\chi^2$. Lower panel: best-fitting model $E^2 f(E)$, with the plot of the reflection components \xillver\ (dotted line) and \relxill\ (dashed lines), the \cloudy\ model for the soft emission lines with associated reflected continuum (solid line), and the \diskbb\ component for the soft excess (dotted lines). The data were rebinned for plotting purposes. \label{fit}}
\end{figure*}

We also tested the possibility that the iron line and the reflection hump are	not directly correlated. We thus included an additional narrow Gaussian line at 6.4 keV, not related to the reflection components. Such a line could originate from Compton-thin gas, which does not produce a significant Compton hump \cite[e.g.,][]{bianchi72132008}. However, the fit is only marginally improved, and the parameters of the reflection components are not significantly altered. Therefore, for the sake of simplicity, we included only \xillver\ and \relxill\ in our model.
\begin{table}
	\begin{center}
		\caption{
			The properties of the iron K emission lines (see text and Figs. \ref{cor1}, \ref{cor2} for the details). $E$ is the energy of the lines (rest-frame), $\sigma$ is the intrinsic line width and EW is the equivalent width. The flux is in units of $10^{-5}$ photons \sqcm\ \pers. Parameters in italics were frozen.
			\label{iron}}
		\begin{tabular}{c c c c}
			\hline \rule{0pt}{2.5ex}
			$E$ (keV) & $\sigma$ (eV) & average flux & average EW (eV) \\ \hline \rule{0pt}{2.5ex}
			\textit{6.4} (narrow) & \textit{0} & 2.42 & 106\\
			\textit{6.4} (broad) & \aer{300}{130}{70} & 2.36 & 102\\
			\textit{7.056} & \textit{0} & 0.6 & 30 \\ \hline
		\end{tabular}
	\end{center}
\end{table}
\paragraph*{The warm absorber and soft lines:} Following Sect. \ref{subsec:rgs}, we included two ionized absorbers and the soft X-ray emission lines found with RGS, modelled with \cloudy. To account for cross-calibration uncertainties between pn and RGS, especially below 0.5 keV, the parameters of the absorbers were left free, but kept constant between the different observations. We found no significant improvement by leaving the warm absorber free to vary between the different observations. The \cloudy\ table for the emission lines was fixed to the RGS best-fitting values, as it was poorly constrained by pn data. This reflected spectrum also adds a weak contribution in the hard X-ray band, because it produces self-consistently an iron line and a reflection hump (see Fig. \ref{fit}).

We also tested a partial covering scenario for the warm absorbers. 
However, we find that if the covering fraction of both components is allowed to vary, it is consistent with 1.
\paragraph*{The soft excess:} Ionized reflection alone does not explain all the observed soft excess in this source, because including only the reflection components leaves strong, positive residuals below 1 keV. We modelled the soft excess phenomenologically with a multicolour disc black-body component, through the \diskbb\ model in \xspec\ \citep{mitsuda1984,makishima1986}. The normalization was left free to vary between different observations, while the inner disc temperature 
was found to be consistent with a common value of around 110 eV, thus it was tied between the different observations. \\ \\

In Fig. \ref{fit} we show the data, residuals and best-fitting model, while all the best-fitting parameters are reported in Table \ref{params}. 
According to our results, the unblurred reflection component (\xillver) is consistent with being constant and due to neutral matter. The reflection strength \rsxill, calculated as the ratio between the 20--40 keV flux of the \xillver\ component and the 20--40 keV primary flux, ranges between $\sim 0.35$ and $\sim 0.6$. This ratio is slightly different from the traditional reflection fraction $\mathcal{R}_f=\Omega/2\pi$, where $\Omega$ is the solid angle subtended by the reflector \cite[][]{pexrav}. We discuss this further in Sect. \ref{sec:discussion}.
The blurred component (\relxill) is due to an ionized disc with $\xi \sim 1000$ \lumcgs\ cm, and with an inner radius $\rin \simeq 40$ \rg. The reflection strength \rsrel, i.e. the ratio between the 20--40 keV flux of \relxill\ and that of the 20--40 keV primary flux, ranges between $\sim 0.15$ and $\sim 0.25$. 
Analogous results are found calculating the ratio between fluxes in a broader energy range, such as 0.1--100 keV \cite[e.g.,][]{wilkins2015}. With this choice, we find similar values for \rsrel, while for \rsxill\ we find 0.15--0.35. However, using the 20--40 keV range is likely more appropriate, as the reflection spectrum in that band is dominated by Compton scattering, only weakly depending on the iron abundance or the ionization state \cite[][]{dauser2016}.
We also estimated the equivalent width of the broad iron line associated with \relxill, finding values consistent with those measured in Sec. \ref{subsec:line}, namely 100--150 eV.

\begin{figure}
	\includegraphics[width=\columnwidth]{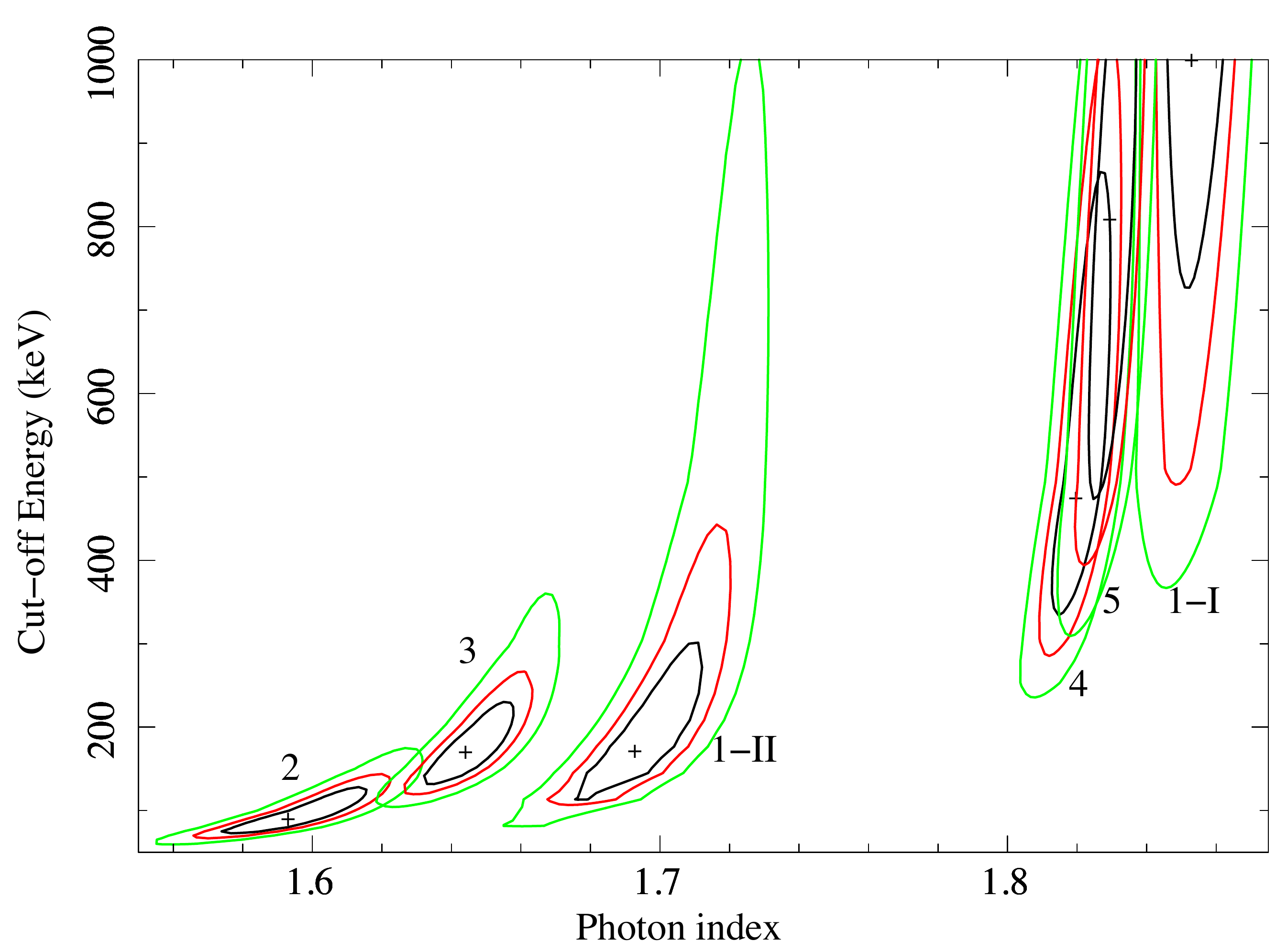}
	\caption{Contour plots of the primary continuum cut-off energy versus photon index for each observation (see also table \ref{params}). Green, red and black lines correspond to 99\%, 90\% and 68\% confidence levels, respectively. }
	\label{gamma_ec}
\end{figure}

Concerning the primary continuum, we find significant variations of the photon index (up to $\Delta \Gamma \simeq 0.2$) and of the high-energy cut-off. The contour plots of the cut-off energy versus photon index are shown in Fig. \ref{gamma_ec}. 
To check the significance of the cut-off energy variability, we repeated the fit keeping the cut-off energy $\cut$ tied between the different observations. We find $\cut=\aerm{420}{400}{130}$ keV, however the fit is significantly worse ($\dchi/\ddof={+}48/{+}5$ with a $p$ value of \expo{3}{-8} from an $F$-test). 
We also performed a fit to the co-added pn and \nus\ spectra (see Table \ref{params}), finding a photon index of \ser{1.84}{0.01} and a cut-off energy \linf{640} keV. More importantly, the parameters are consistent with those that were kept constant in the simultaneous fit of the single observations.

\begin{figure}
	\includegraphics[width=\columnwidth]{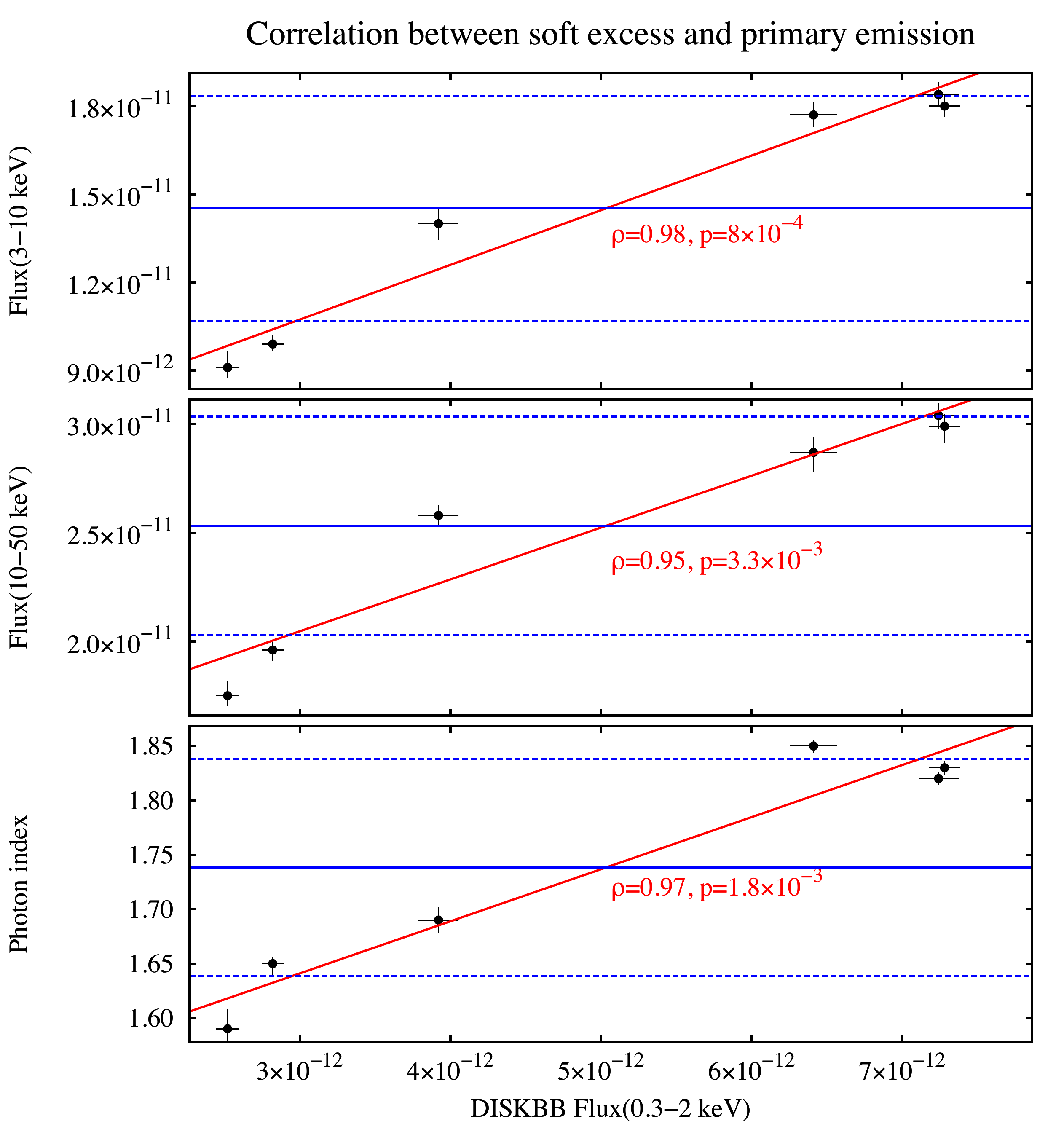}
	\caption{Three parameters of the primary power law plotted against the flux of the soft excess (calculated as the 0.3--2 keV flux of the \diskbb\ component, see Sect. \ref{subsec:broad}). Upper panel: primary flux in the 3--10 keV range. Middle panel: primary flux in the 10--50 keV range. Lower panel: primary photon index. All fluxes are in units of \fluxcgs. The blue solid lines represent the mean value for each parameter, while the blue dashed lines represent the standard deviation. The red lines represent linear fits to the data. }
	\label{se_flux}
\end{figure}

A multicolour disc black-body with an inner temperature of $\sim 110$ eV is found to describe well the soft excess seen on top of the primary power law and of the ionized reflection component.
We note that, if the primary power law is due to Comptonization, it should have a low-energy roll-over dependent on the temperature of the seed photons, which could enhance the soft excess. However, if the seed photons are in the optical/UV band, the effect is negligible when fitting the data above 0.3 keV (see also the discussion in Sect. \ref{sec:discussion}).
We also found a significant correlation between the soft excess flux and the primary emission.
We show in Fig. \ref{se_flux} the 3--10 keV flux of the primary power law, the 10--50 keV flux and the photon index, plotted against the \diskbb\ flux in the 0.3--2 keV band. The correlation between the \diskbb\ flux and the soft primary flux has a Pearson's coefficient of 0.98 and a $p$ value of \expo{8}{-4}. For the hard band, the Pearson's coefficient is 0.95, with a $p$ value of \expo{3.3}{-3}. For the photon index, the Pearson's coefficient is 0.97, with a $p$ value of \expo{1.8}{-3}.
A detailed modelling of the soft excess is beyond the scope of this paper, and it is deferred to a following paper.
\section{Discussion and conclusions}\label{sec:discussion}
We reported results based on a joint \xmm\ and \nus\ monitoring of NGC 4593, aimed at studying the broad-band variability of this AGN in detail.

The source shows a remarkable variability, both in flux and in spectral shape, on time-scales of a few days and down to a few ks. While flux variability is significant at all energies, the spectral variability is mostly observed below 10 keV. Moreover, the spectrum is softer when the source is brighter, as generally found in luminous Seyfert galaxies \cite[e.g.,][]{markowitz2003,sobolewska&papadakis,soldi2014}. If the X-ray emission is produced by Comptonization, the relation between observed flux and photon index can be understood as an effect of that process. The spectral shape primarily depends on the optical depth and temperature of the Comptonizing corona, albeit in a non-trivial way \cite[e.g.,][]{sunyaev&titarchuk1980}. The photon index can be also related to the Compton amplification ratio, i.e. the ratio between the total corona luminosity and the soft luminosity that enters and cools the corona. A higher amplification ratio corresponds to a harder spectrum \cite[e.g.,][]{haardt&maraschi1993}. In the limiting case of a ``passive'' disc, which is not intrinsically radiative but rather reprocesses the coronal emission, the amplification ratio is fixed by geometry alone \cite[e.g.,][]{haardt&maraschi1991,stern1995}. Then, in steady states (i.e. the system is in radiative equilibrium), the coronal optical depth and temperature adjust to keep the amplification ratio constant. 
But the amplification ratio can change following geometry variations of the accretion flow. For example, the inner radius of the disc may be different from the innermost stable circular orbit (as suggested by the fit with the
relativistic reflection) and change in time\footnote{We do not detect significant variations of the inner radius of the \relxill\ reflection component. However, given the uncertainty, we cannot exclude variations by up to a factor of 2 (from $\sim 25$ to $\sim 50$ \rg).}, causing a decrease of the soft luminosity supplied by the disc. This, in turn, may increase the amplification ratio, leading to a harder spectrum.
This kind of behaviour is especially apparent during obs. 1, where a significant flux drop (a factor of $\sim 2$ in the 0.5--2 keV count rate) is accompanied by a significant hardening of the spectrum ($\Delta \Gamma \simeq 0.15$). 

However, we also find a significantly variable high-energy cut-off (see Fig. \ref{gamma_ec}). In particular, the lowest cut-off is seen during obs. 2, where $\cut =\aerm{90}{40}{20}$ keV. This corresponds to the hardest spectrum, with $\Gamma \simeq 1.6$. For the softest spectra, like obs. 1-I and obs. 5, we only find a lower limit on $\cut$ of several hundreds of keV. Since the cut-off energy is generally related to the coronal temperature \cite[$\cut \simeq 2\mbox{-}3 \, kT$, e.g.][]{poptestingcompt2001}, our results might suggest that this temperature undergoes huge variations (from a few tens up to a few hundreds of keV) during a few days. However, it should be noted that the turnover of a Comptonized spectrum is much sharper than a simple exponential cut-off \cite[e.g.,][]{zdziarski2003}. Then, fitting the data with a cut-off power law provides only a phenomenological description that does not allow for an unambiguous physical interpretation \cite[e.g.,][]{poptestingcompt2001}. 

We confirm the presence of a warm absorber affecting the soft X-ray spectrum. \cite{ebrero_4593}, using a 160 ks \chandra\ data set, found evidence for a multi-phase ionized outflow with four distinct degrees of ionization, namely $\log \xi = 1.0, 1.7, 2.4$ and $3.0$. The low-ionization components were found to have lower velocities (300-400 \kms) than the high-ionization components (up to $\sim 1000$ \kms). According to our results, a two-phase ionized outflow provides an adequate characterization of this gas, both for RGS and pn spectra. These two components have different degrees of ionization, namely $\log \xi \simeq 2.2\mbox{-}2.5$ and $0.1\mbox{-}0.4$, and different column densities, $\nh \simeq (2\mbox{-}7) \times 10^{21}$ and $(0.8\mbox{-}1.2) \times 10^{21}$ \sqcm, with uncertainties due to the different estimates based on pn and RGS data. The two components are also kinematically distinct, with the high-ionization component outflowing faster ($v \simeq -900$ km~\pers) than the low-ionization one ($v \simeq -300$ km~\pers). 
We can derive a rough upper limit on the distance of such components in the following way \cite[see also][]{ebrero_4593}. By definition, $\xi = L_{\textrm{ion}}/ n R^2$, where $L_{\textrm{ion}}$ is the ionizing luminosity in the 1--1000 Ryd range, $n$ is the hydrogen gas density and $R$ the distance. If we assume that the gas is concentrated within a layer with thickness $\Delta r \leq R$, we can write $R \leq L_{\textrm{ion}} / \nh \xi$, where $\nh = n \Delta r$. The average ionizing luminosity that we obtain from our best-fitting model is $\sim 1 \times 10^{43}$ \lumcgs. Then, from the measurements of $\nh$ and $\xi$ we get an upper limit on $R$. For the high-ionization component we estimate $R \lesssim 3$ pc, while for the low-ionization component $R \lesssim  1.5$ kpc. 
The outflows are thus consistent with originating at different distances from the nucleus, as found by \cite{ebrero_4593}, with the high-ionization and high-velocity component possibly being launched closer to the central source. 

We find evidence for a cold, narrow \fek\ line which is consistent with being constant during the monitoring. A broad component is also detected, and is consistent with originating at a few tens of gravitational radii from the black hole. The \fek\ line is accompanied by a moderate Compton hump (the total reflection strength is approximately 0.5--0.8). All these features are well modelled by including two different reflection components, both requiring an iron overabundance of 2-3.
One component (\xillver) is responsible for the narrow core of the \fek\ line, and is consistent with arising from a slab of neutral, Compton-thick material. This component is also consistent with being constant, despite the observed strong variability of the primary continuum. The variability of the primary emission can be diluted in a distant reflector, if the light-crossing time across the reflecting region is more than the variability time-scale of the primary continuum. In this case, the reflector responds to a time-averaged illumination from the primary source. Therefore, the reflecting material should lie at least a few light days away from the primary source, and it can be interpreted as the outer part of the accretion disc.
The reflection strength \rsxill, namely the ratio between the reflected and the primary flux in the 20--40 keV band, is in the range 0.35--0.6.
According to \cite{dauser2016}, the reflection strength is likely an underestimate of the reflection fraction $\mathcal{R}_f$, defined as the ratio between the primary intensity illuminating the disc and the intensity reaching the observer. The difference is that $\mathcal{R}_s$ depends on the inclination of the reflector, while $\mathcal{R}_f$ does not. For an isotropic source, $\mathcal{R}_f=\Omega / 2\pi$, where $\Omega$ is the solid angle covered by the reflector as seen from the primary source. \cite{dauser2016} report $\mathcal{R}_s \simeq 0.7$ when $\mathcal{R}_f$ is set to 1 in \xillver, for an inclination angle of 50 deg. Therefore, in our case, we estimate \rfxill\ to be in the range 0.5--0.85. This would correspond to a solid angle between $\pi$ and $1.7\pi$. However, there are a few caveats. First, if the reflecting material is more distant (e.g. a torus at pc scales), the reflection fraction might not be closely related to the viewing angle \cite[e.g.,][]{malzac&petrucci2002}. Furthermore, the X-ray emitting corona could be outflowing or inflowing rather than being static \cite[][]{belo1999,malzac2001}. In this case, the reflection fraction would depend also on the velocity of the corona, because the relativistic beaming would result in an anisotropic illumination \cite[][]{malzac2001}. 

The second reflection component (\relxill) is responsible for the broad \fek\ line, and is consistent with originating from the inner part of the accretion disc, down to $\sim 40$ \rg. 
Given the black hole mass of $\sim 1 \times 10^7$ solar masses, a distance of $40$ \rg\ corresponds to a light-travel time of $\sim 2$ ks. Indeed, this reflection component appears to respond to the variability of the primary continuum (see Table \ref{params} and Fig. \ref{cor2} from the phenomenological fit of the broad \fek\ line with a Gaussian component).
A more detailed discussion is deferred to a forthcoming paper, where a timing analysis will be performed (De Marco et al., in prep.).
The reflection strength \rsrel\ is $\sim 0.2$, and the effects of light bending in \relxill\ may complicate the geometrical interpretation \cite[][]{dauser2016}. However, given the relatively large inner disc radius, we do not expect a strong modification of the reflection fraction \cite[e.g.,][]{dauser2014}. Using the same relations as for \xillver, we find a covering angle $\Omega \simeq \pi$. 
Finally, our results suggest that a slightly larger \rsrel\ corresponds to the low-flux states, i.e. to a flatter $\Gamma$, whereas a correlation between the photon index and the reflection fraction is usually found in AGNs \cite[e.g.][]{ZLS99,malzac&petrucci2002}. According to \cite{ZLS99}, a correlation can be explained if the reflecting medium is also the source of seed soft photons that get Comptonized in the hot corona. In our case, the lack of such a correlation and the presence of a strong soft excess suggest that the source of soft photons and the reflecting medium are not one and the same.

The soft X-ray excess
is phenomenologically well described by a variable multicolour disc black-body. Physically, the soft excess could be produced by a warm corona upscattering the optical/UV photons from the accretion disc \cite[see, e.g.,][]{mag_5548,pop2013mrk509}. This warm corona could be the upper layer of the disc \cite[e.g.,][]{rozanska2015}.
The observed correlation between the flux of the soft excess and that of the primary power law (Fig. \ref{se_flux}) could indicate that these two components arise from the Comptonization of the same seed photons. Then, the observed correlation between the soft excess and the primary photon index can be seen as a simple consequence of the flux correlation and of the softer-when-brighter behaviour of the primary continuum. However, the actual scenario could be more complex. Assuming that the soft excess is produced by a warm corona, part of this emission (depending on geometry) could be in turn upscattered in the hot corona and contribute to its Compton cooling. In this case, a larger flux from the warm corona would imply a more efficient cooling of the hot corona, thus a softer X-ray emission. These results motivate a further, detailed investigation using realistic Comptonization models, which will be the subject of a forthcoming paper.
\begin{table*}
	\begin{center}
		\caption{Best-fitting parameters of the broad-band (0.3--80 keV) phenomenological model described in Sect. \ref{subsec:broad}: {\sc WA1*WA2*(soft lines + cutoffpl + diskbb + relxill + xillver)}. In the second column we report the fit parameters that were kept constant, i.e. they were constrained by all observations. 
		In the third to eighth column,
		we report the fit parameters that were free to vary for each observation. 
		In the last column, we report the best-fitting parameters for the summed spectra.
		Parameters in italics were frozen due to poor constraints. \label{params}
		}
		\begin{tabular}{l c c c c c c c c}
			\hline \rule{0pt}{2.5ex} & all obs. & obs. 1-I & obs. 1-II & obs. 2 & obs. 3 & obs. 4 & obs. 5 & summed\\
			\hline \hline \rule{0pt}{2.5ex} 
			$\log \xione$ & \ser{2.71}{0.05} &&&&&&& \aer{ 2.70 }{ 0.09 }{ 0.05 } 
			\\ \rule{0pt}{2.5ex} 
			$\log \vturbone$ & \linf{2.2} &&&&&&& $\mathit{2.5}$
			\\ \rule{0pt}{2.5ex} 
			$\log \nhone$ & \ser{21.4}{0.1} &&&&&&& \ser{21.4}{0.1}  \\ \rule{0pt}{2.5ex}
			$\log \xitwo$ & $\mathit{0.1}$ &&&&&&&$\mathit{0.1}$
			\\ \rule{0pt}{2.5ex} 
			$\log \vturbtwo$ & \aer{2.01}{0.01}{0.05} &&&&&&& \aer{ 2.2 }{ 0.2 }{ 0.4 }
			\\ \rule{0pt}{2.5ex} 
			$\log \nhtwo$ & \aer{21.09}{0.01}{0.07} &&&&&&& \aer{ 21.13 }{ 0.02 }{ 0.03 } 
			\\  \hline \rule{0pt}{2.5ex} 
			$\Gamma$ && \ser{1.85}{0.01} 
			& \ser{1.69}{0.02}  & \aer{1.59}{0.03}{0.02} &\aer{1.65}{0.01}{0.02}  & \ser{1.82}{0.01}  & \ser{1.83}{0.01} & \ser{1.84}{0.01} \\ \rule{0pt}{2.5ex} 
			$\cut$ (keV)& & \linf{700} 
			& \aer{170}{160}{60}  & \aer{90}{40}{20}  & \aer{170}{70}{40}  & \aer{470}{430}{150}  & \linf{450} & \linf{640}  \\ \rule{0pt}{2.5ex} 
			$N_{\textsc{pow}} (\times 10^{-3})$ &&  \aer{7.1}{0.2}{0.1} 
			& \ser{4.4}{0.2}  & \aer{2.5}{0.2}{0.1}  & \ser{2.9}{0.1}  & \ser{7.1}{0.2}  & \aer{7.0}{0.2}{0.1} &\ser{5.4}{0.1}   \\ \hline \rule{0pt}{2.5ex} 
			$kT_{\textsc{diskbb}}$ (eV) & \ser{110}{3}
			&  &   &  &   &   & & \aer{105 }{4}{3} \\ \rule{0pt}{2.5ex} 
			$N_{\textsc{diskbb}} (\times 10^3)$  && \aer{6.8}{0.2}{0.3}
			& \aer{4.1}{0.3}{0.2}  & \aer{2.7}{0.1}{0.2}  & \aer{2.9}{0.2}{0.1}  & \aer{7.6}{0.3}{0.2}  & \ser{7.7}{0.2}& \ser{7.5}{1.6}{1. }   \\ \hline \rule{0pt}{2.5ex} 
			$\log \xi_{\textsc{relxill}}$ & \aer{3.00}{0.01}{0.13}
			& &   &   &   & & & \aer{ 2.93}{ 0.08 }{ 0.11 } \\ \rule{0pt}{2.5ex} 
			$\rin$ (\rg) & \ser{ 40 }{ 15 }&&&&&&&\aer{ 40 }{ 15 }{ 30 }  \\ \rule{0pt}{2.5ex} 
			$N_{\textsc{relxill}} (\times 10^{-5})$& & \ser{3.1}{0.7}
			& \ser{2.5}{0.5}  & \ser{2.4}{0.3}  & \ser{2.6}{0.3}  & \aer{4.6}{0.6}{0.5}  & \aer{4.2}{0.4}{0.5} & \aer{ 3.3 }{ 0.7 }{ 0.5 }
			\\ \hline \rule{0pt}{2.5ex} 
			$N_{\textsc{xillver}} (\times 10^{-5})$  & \ser{2.5}{0.3}&&&&&&& \aer{ 3.1 }{ 0.3 }{ 0.4 } 
			\\ \rule{0pt}{2.5ex} 
			$\afe$  & \aer{2.6}{0.2}{0.4}&&&&&&& \ser{ 2.0 }{ 0.3}
			\\ \hline \rule{0pt}{2.5ex} 
			$\rchisq$&$2506/2299$&&&&&&&2401/2264 \\ \rule{0pt}{2.5ex} 
			individual $\rchisq$& & $383/362$ & $389/340$ & $380/376$ & $404/387$& $491/426$ & $459/421$&  \\ \hline
		\end{tabular}
	\end{center}
\end{table*} 
\section*{Acknowledgements}
We thank the anonymous referee for helpful comments and suggestions that improved the paper.\\
This work is based on observations obtained with: the \nus\ mission, a project led by the California Institute of Technology, managed by the Jet Propulsion Laboratory and funded by NASA; \xmm, an ESA science mission with instruments and contributions directly funded by ESA Member States and the USA (NASA). This research has made use of data, software and/or web tools obtained from NASA's High Energy Astrophysics Science Archive Research Center (HEASARC), a service of Goddard Space Flight Center and the Smithsonian Astrophysical Observatory. 
FU, POP, GM, SB, MC, ADR and JM 
acknowledge support from the french-italian International Project of Scientific Collaboration: PICS-INAF project number 181542. FU, POP acknowledge support from CNES. FU acknowledges support from Universit\'e Franco-Italienne (Vinci PhD fellowship). FU, GM acknowledges financial support from the Italian Space Agency under grant ASI/INAF I/037/12/0-011/13. POP acknowledges financial support from the Programme National Hautes Energies. SB, MC and ADR acknowledge financial support from the Italian Space Agency under grant ASI-INAF I/037/12/P1. 
GP ackowledges support by the Bundesministerium f\"ur Wirtschaft und Technologie/Deutsches Zentrum f\"ur Luft und Raumfahrt (BMWI/DLR, FKZ 50 OR 1408) and the
Max Planck Society.
\bibliographystyle{mnras}
\bibliography{mybib.bib}
\end{document}